\documentclass[useAMS,usenatbib,a4paper]{mn2e}
\usepackage{graphicx}
\usepackage{amsmath}
\usepackage{amssymb}
\usepackage{color}
\def\OJ{OJ 287}

\def\la{\langle }

\newcommand{\be}{\begin{equation}}
\newcommand{\ee}{\end{equation}}
\newcommand{\bea}{\begin{eqnarray}}
\newcommand{\eea}{\end{eqnarray}}
\newcommand{\bg}{\begin{gather}}
\newcommand{\eg}{\end{gather}}
\newcommand{\bseq}{\begin{subequations}}
\newcommand{\eseq}{\end{subequations}}

\def\theenumi{(\alph{enumi})}

\def\gr{$\gamma$-ray}

\begin{document}

\title{Fast variability of $\gamma$-ray emission from supermassive black hole binary OJ 287}

\author[Neronov and Vovk]{A.~Neronov$^{1}$, Ie.~Vovk$^{1}$\\ 
$^{1}$ISDC Data Center for Astrophysics, Geneva Observatory, Chemin d'\'Ecogia 16, 1290
Versoix, Switzerland}
\date{Received $<$date$>$  ; in original form  $<$date$>$ }
\pagerange{\pageref{firstpage}--\pageref{lastpage}} %\pubyear{2005}
\maketitle
\label{firstpage}

\begin{abstract}
We report the discovery of fast variability of $\gamma$-ray flares from blazar OJ 287. This blazar is known to be powered by binary system of supermassive black holes. The observed variability time scale $T_{\rm var}\lesssim 3-10$~hr is much shorter than the light crossing time of more massive ($1.8\times 10^{10}M_\odot$) black hole and is comparable to the light crossing time of the less massive   ($1.3\times 10^8M_\odot$) black hole. This indicates that  \gr\ emission is produced by relativistic jet ejected by the black hole of smaller mass. Detection of \gr s with energies in excess of 10 GeV during the fast variable flares  constrains the Doppler factor of the jet to be larger than  4. Possibility of the study of orbital modulation of emission from relativistic jet makes \OJ\ a unique laboratory for the study of the mechanism(s) of formation of jets by black holes, in particular, of the response of the jet parameters to the changes of the parameters of the medium from which the black hole accretes and into which the jet expands. 
\end{abstract}


%%%%%%%%%%%%%%%%%%%%%%%%%%%%%%%%%%%%%%%%%%%%
\section{Introduction}
%%%%%%%%%%%%%%%%%%%%%%%%%%%%%%%%%%%%%%%%%%%%

Although large-scale jets ejected by Active Galactic Nuclei (AGN) were discovered almost a century ago, origin of this phenomenon remains obscure up to present days (see \citet{jet_review} for a recent review). It might be that the jets are accelerated via magneto-centrifugal force along twisted magnetic field lines above accretion disk around the black hole \citep{blandford-payne}. Otherwise, an outflow can be created via Blandford-Znajek mechanism of electromagnetic power extraction from a rotating BH, similar to the mechanism responsible for the generation of relativistic pulsar winds \citep{blandford-znajek}. 

Blazar \OJ\ ($z=0.306$ \citep{stickel89}) provides a unique laboratory for the study of the mechanism of AGN activity, because this is one of the few AGN known to host binary black hole system \citep{lehto96,valtonen09}. In this system, a lighter black hole of the mass $M_{\rm BH1}\simeq 1.3\times 10^8M_\odot$ orbits a heavier black hole of the mass $M_{\rm BH2}\simeq 1.8\times 10^{10}M_\odot$ with a period $P_{\rm orb}\simeq 11.65$~yr \citep{valtonen09,sillanpaa88}. Separation of the components of the system at periastron is just about 10 Schwarzschild radii of the heavier black hole, so that the orbital motion is strongly affected by relativistic gravity effects \citep{OJ_nature}. 

\OJ\ is known to belong to the BL Lac sub-class of AGNi, which means that it emits a relativistic jet whose direction is aligned with the line of sight.  It is not clear a priori, which of the black holes ejects the observed relativistic jet. Following a naive argument, which does not take into account the relativistic beaming of the jet emission, one would assume that the observed relativistic jet is the one ejected by the heavier black hole, simply because the bigger black hole accretes more matter and, therefore, could produce more powerful jet. Most of the existing studies of multi-wavelength blazar activity of \OJ\ adopt this assumption (see e.g. \citet{valtonen09}). However, relativistic jets in  BL Lacs are known to Doppler factors $\delta\gg 1$. This results in boosting the apparent luminosity of the jets by a factor $\delta^4$, when the jets are viewed face-on. Thus, if the less powerful jet emitted by the smaller black hole is aligned with the line of sight, while the jet from the larger black hole is not, the jet from the smaller black hole might give dominant contribution to the source flux.

In what follows we show that variability properties of \gr\ emission from the source indicate that the  
relativistically beamed emission comes from the jet produced by the smaller black hole.  Independently of the value of the Doppler factor of the jet $\delta$, the shortest observed variability time scale $\Delta T_{min}$ imposes a constraint on the size of the jet's "central engine", $R_{\rm CE}\lesssim c\Delta T$ \citep{celotti98,neronov09}. In the case of \OJ,  $R_{\rm CE}$ turns out to be much smaller than the Schwarzschild radius of the more massive black hole, but compatible with the size of the smaller mass black hole. 

Observation of \gr\ emission from the base of the jet of the smaller black hole in the system makes \OJ\ a unique laboratory for the study of mechanisms of jet production. Regular orbital modulation of the physical parameters of the ambient medium around the $1.3\times 10^8M_\odot$ black hole provides a unique possibility to study the response of the jet to the changes of the properties of accretion flow and of external medium in which the jet propagates.  In this respect, the \OJ\ system provides a scaled-up analog of Galactic \gr -loud binaries, in which orbital modulation of  \gr\ emission enables a study of response of relativistic outflow from a compact object (a neutron star or a black hole) to the changes of the properties of external medium (stellar wind and radiation field of companion star) (see e.g. \cite{zdz10}).

%%%%%%%%%%%%%%%%%%%%%%%%%%%%%%%%%%%%%%%%%%%%
\section{{\it Fermi} observations}
%%%%%%%%%%%%%%%%%%%%%%%%%%%%%%%%%%%%%%%%%%%%

%%%%%%%%%%%%%%%%%%%%%%%%%%%%%%%%%%%%%%%%%%%%
\begin{figure}
\includegraphics[width=\columnwidth]{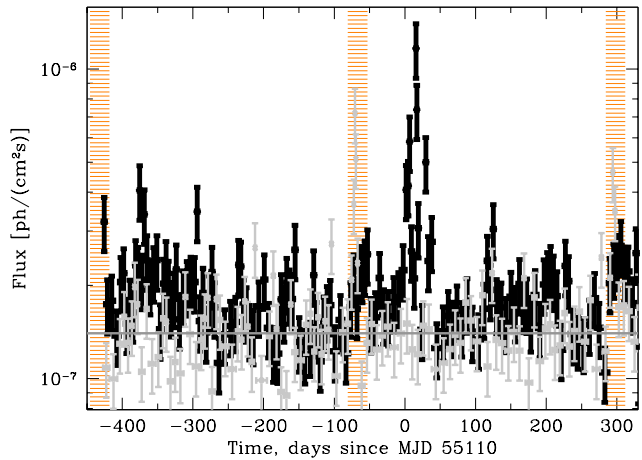}
\caption{August 2008 -- January 2010 lightcurve of \OJ\  (black) and background region (grey). Orange bands mark the periods of passage of the Sun through the observation region. }
\label{fig:lcurve_longterm}
\end{figure}
%%%%%%%%%%%%%%%%%%%%%%%%%%%%%%%%%%%%%%%%%%%%

%%%%%%%%%%%%%%%%%%%%%%%%%%%%%%%%%%%%%%%%%%%%
\begin{figure}
\includegraphics[width=\linewidth]{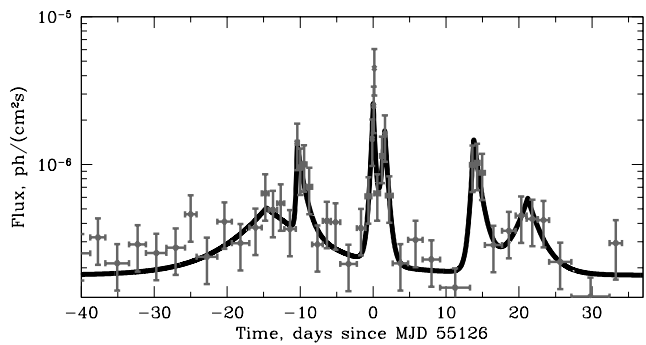}
\caption{Lightcurve of October-November 2009 flare of \OJ\ in $E>0.1$~GeV energy band, binned in time bins with S/N ratio equal to 3.  Black curve shows model fit to the lightcurve, Eq. (\ref{eq:model}) with parameters given in Table \ref{tab:model}.}
\label{fig:lcurve}
\end{figure}
%%%%%%%%%%%%%%%%%%%%%%%%%%%%%%%%%%%%%%%%%%%%

In order to study variability of the \gr\ signal during the flaring activity, we have processed publicly available data of the LAT instrument, using the {\it Fermi} Science Tools provided by the {\it Fermi} Science Support Centre. The data were selected using {\it gtselect} tool. The lightcurves were produced with the help of {\it gtbin} and {\it gtexposure} tools as it is explained in the {\it Fermi} data Analysis Threads \footnote{\tt http://fermi.gsfc.nasa.gov/ssc/data/analysis/scitools/}. 
%%%%%%%%%%%%%%%%%%%%%%%%%%%%%%%%%%%%%%%%%%%%
\begin{figure}
\includegraphics[width=\linewidth]{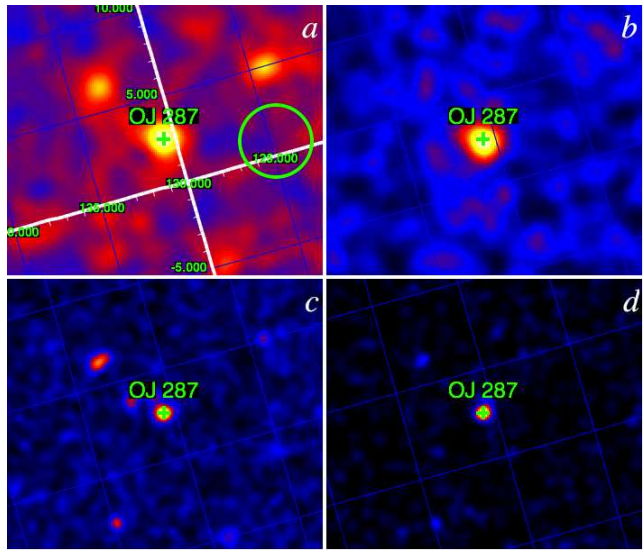}
\caption{Images of sky region around \OJ. Panels {\it a} and {\it c} show 0.3-1~GeV and 1-300~GeV band images for the entire observation period. Panels {\it b} and {\it d} show 0.3-1~GeV and 1-300~GeV band images during the flaring activity of \OJ. Coordinate grid show ecliptic coordinates. Green circle shows the region from which the background lightcurve was extracted. Images in the 0.3-1~GeV band are smoothed with a gaussian of the width $1^\circ$. Images in the 1-300~GeV band are smoothed with the Gaussian with $0.5^\circ$ width.  }
\label{fig:image}
\end{figure}
%%%%%%%%%%%%%%%%%%%%%%%%%%%%%%%%%%%%%%%%%%%%

Fig. \ref{fig:lcurve_longterm} shows the longterm lightcurve of the source at the energies above 0.1~GeV over the August 2008 -- September  2010 period binned to achieve signal to noise ratio S/N=5 per time bin. Photons were collected from the circle $2^\circ$ in radius and centered on \OJ\ . The flaring activity period, which started on October 5 2009, could be readily identified. The detailed lightcurve of the flaring period shown in Fig. \ref{fig:lcurve} reveals several well separated flares. The brightest flare which happened on October 22, 2009 was reported by \citet{fermi_atel}. Followup observations of October 22 flare by {\it Swift}/XRT have revealed an increase of the X-ray flux accompanying the \gr\ flare \citep{swift}.

The direction toward \OJ\ is situated close to the plane of Ecliptic. The lightcurve of the source could be affected  by the passage of the Sun through the field of view of the telescope. The periods of the 
Sun passage within the region of the radius $15^\circ$ centered on \OJ\ (the sky region chosen for the analysis) are shown by the orange-shaded strips in Fig. \ref{fig:lcurve_longterm}. Possible effect of the Sun passage on the source lightcurve is most clearly seen in the lightcurve collected from a region of the radius $2^\circ$ displaced by $\sim 5^\circ$ from the position of \OJ\  to the position RA=127.11 DEC=19.89. No source is detected at this position, so the lightcurve collected from this region (shown by grey data points in Fig. \ref{fig:lcurve_longterm}) could be considered as a measure of the diffuse background level close to the position of \OJ. One could notice a flare in the background lightcurve, associated with the Sun passage close to the background extraction region.  The effect of the Sun passage on the lightcurve of \OJ\ is less pronounced (see Fig. \ref{fig:lcurve_longterm}), because the source is situated higher above the Ecliptic plane, compared to the background region.

Analysis of the images of \gr\ sky around \OJ\ show that, apart from \OJ, the region of the radius $15^\circ$ centered on \OJ\ contains several other sources (most clearly visible in the 1-300~GeV band image in the panel {\it c} of Fig. \ref{fig:image}). However, these sources are rather weak, so that they are not detected on one month exposure time scale corresponding to the duration of the flaring activity of \OJ (see the right panel of Fig. \ref{fig:image}). Absence of strong sources near \OJ\ ensures that the observed variability of the signal from the source during the flaring period is not affected by possible variable emission from a nearby source for which the tail of the point spread function overlaps with that of \OJ.

The $E>0.1$~GeV lightcurve of the flare consists of several well separated pulses with rather sharp rise and decay.  To find the rise and decay times we have fitted the lightcurve with a phenomenological model of a sum of exponentially rising and decaying pulses  
\begin{equation}
F(t)=B+\sum_{k=1}^3\left\{
\begin{array}{ll}
A_k\exp\left((t-t_k)/t_{rk}\right), & t<t_k\\
A_k\exp\left(-(t-t_k)/t_{dk}\right), & t>t_k
\end{array}
\right.
\label{eq:model}
\end{equation}
where $B=const$ is the background level. The background level was found from the circle of the radius $2^\circ$ displaced by 5 degrees from the source position. Parameters of the model function (\ref{eq:model}), derived from the fitting, are given in Table \ref{tab:model}.

%%%%%%%%%%%%%%%%%%%%%%%%%%%%%%%%%%%%%%%%%%%%
\begin{table}
\begin{tabular}{c|cccc}
\hline
k&$t_k$&$t_{rk}$&$t_{dk}$&$A_k$\\
&d& d&d&$\times 10^{-6}$~cm$^{-2}s^{-1}$\\
\hline
1&$1.4\pm 0.5$& $5.0_{-2.5}^{+10.0}$&$7.0^{+4.0}_{-3.5}$&$0.33_{-0.08}^{+0.07}$\\
2&$5.6_{-0.3}^{+0.2}$& $0.15_{-0.14}^{+0.37}$&$1.0^{+0.6}_{-0.4}$&$1.0_{-0.3}^{+0.4}$\\
3&$16.02_{+0.07}^{-0.05}$& $0.33_{-0.09}^{+0.19}$&$0.30_{-0.1}^{+0.15}$&$2.3^{+0.6}_{-0.6}$\\
4&$17.6_{-0.15}^{+0.18}$ & $0.5^{+0.4}_{-0.25}$&$0.46^{+0.25}_{-0.18}$&$1.5_{-0.5}^{+0.5}$\\
5&$29.8_{-0.3}^{+0.4}$& $0.4^{+0.3}_{-0.23}$&$1.0^{+0.5}_{-0.4}$&$1.3_{+0.3}^{-0.3}$\\
6&$37.2_{-1.0}^{+1.3}$& $2.0^{+1.8}_{-1.0}$&$1.9^{-0.9}_{-2.2}$&$0.41_{+0.12}^{-0.14}$\\
\hline
\end{tabular}
\caption{Parameters of the model  fit (see Eq. (\ref{eq:model})) to the lightcurve of the \gr\ flare of \OJ \  together with their $68\%$ confidence ranges. }
\label{tab:model}
\end{table}
%%%%%%%%%%%%%%%%%%%%%%%%%%%%%%%%%%%%%%%%%%%%

\begin{figure}
\includegraphics[width=\linewidth]{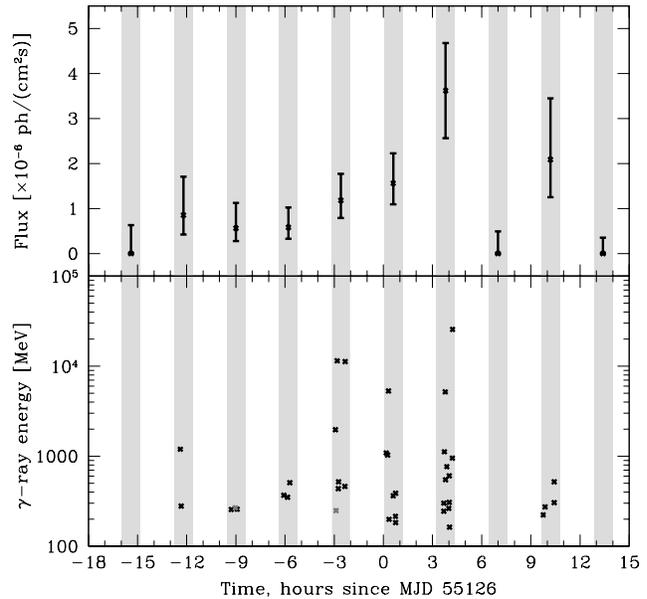}
\caption{Upper panel: lightcurve of brightest episode of October-November 2009 flare of \OJ\ in $E>0.1$~GeV energy band. Lower panel: energies and arrival times of \gr\ from the source (black points) and from the background region (grey points). Vertical grey strips show the periods when the source was in the field of view of LAT telescope.  }
\label{fig:lcurve_zoom}
\end{figure}

One can see from Fig. \ref{fig:lcurve} and Table \ref{tab:model} that brightest flares are characterized by the rather sharp rises and decays, with the rise/decay times of several hours. At this time scales measurement of the rise/decay times of the flares is complicated by the fact that {\it Fermi}/LAT telescope observes a given patch of the sky once in $3.2$ hours (once per two rotation periods of $\simeq 96$~min). This is clear from Fig. \ref{fig:lcurve_zoom} in which lightcurve for the time interval of several hours around the brightest flare from the source is shown in more details. The upper panel of the figure shows the $E\ge 0.1$~GeV lightcurve of the source. The lower panel shows the energies of photons collected from the circle of the radius $2^\circ$ centered on the source (black points) and photons collected from a background extraction circle displaced of the same radius displaced by 5 degrees from the source position. 
Photons from the source and background regions come only within periodic time intervals spaced by 3.2~hr marked by vertical grey strips in the two panels of the Figure. It is clear that {\it Fermi}/LAT pointing pattern does not allow to constrain the rise/decay time of the flares to better than 3.2~hr. The apparently abrupt end of the flare (12 photons from the source are detected between $t=$MJD55126.0$+3$~hr and $t=$MJD55126.0$+4$~hr, while no photon is detected within the subsequent observation period MJD55126.0$+6.2$~hr$<t<$MJD55126.0$+7.2$~hr) limits the decay time of the flare to be less than $3.2$~hr. Assuming that the source flux did not change in the time interval following the peak of the flare, one could estimate the chance probability of detecting zero photons in this time bin to be  $7\times 10^{-4}$, taking into account the exposures in the two adjacent time intervals with maximal and zero count rates, $3.3\times 10^6$~cm$^2$s and $2\times 10^6$~cm$^2$s, respectively. 

To summarize, both fitting of the lightcurve profile with a phenomenological exponential rise / exponential decay model and and direct photon counting in individual  1 hr long {\it Fermi} exposures indicate that the flux from the source is significantly variable on time scales shorter or comparable to 3-10~hr. This fact has important implications for the physical model of the origin of \gr\ emission from the \OJ\ system.

%%%%%%%%%%%%%%%%%%%%%%%%%%%%%%%%%%%%%%%%%%%%
%%%%%%%%%%%%%%%%%%%%%%%%%%%%%%%%%%%%%%%%%%%%

%%%%%%%%%%%%%%%%%%%%%%%%%%%%%%%%%%%%%%%%%%%%
\section{Origin of relativistic jet in \OJ}
%%%%%%%%%%%%%%%%%%%%%%%%%%%%%%%%%%%%%%%%%%%%

Constraint on the variability time scale of the \gr\ flares $T_{\rm var}=\mbox{min}(t_{rk},t_{dk})\le 3.2$~hr, derived above, enables to identify the \gr\ emission site within the binary black hole system of  \OJ.

It is commonly accepted that \gr\ emitting jets are generated by the AGN "central engines", the supermassive black holes, on the distance scales of the order of gravitational radius of supermassive black hole
\begin{equation}
\label{Rg}
R_g=\frac{G_NM_{\rm BH}}{c^2}=2\times 10^{11}\left[\frac{M_{\rm BH}}{1.3\times 10^{8}M_\odot}\right]\mbox{ cm,}
\end{equation}
where $M_{\rm BH}$ is the black hole mass. 
Minimal variability time scale of electromagnetic emission originating from the AGN central engine is expected to be not shorter than the light-crossing time of the supermassive black hole,
\begin{eqnarray}
\label{lc}
T_{\rm lc}&=&(1+z)2R_{\rm BH}/c=2(1+z)\left(R_g+\sqrt{R_g^2-a^2}\right)/c\\
&\simeq&\left\{
\begin{array}{ll}
0.5\left[M_{\rm BH}/1.3\times 10^{8}M_\odot\right]\mbox{ hr,}&a=R_g\\
0.9\left[M_{\rm BH}/1.3\times 10^{8}M_\odot\right]\mbox{ hr,}&a=0
\end{array}
\right.\nonumber
\end{eqnarray} 
where $R_{\rm BH}$ is the size of the black hole horizon and $0<a<R_g$ is black hole rotation moment per unit mass. Variability of X-ray emission at time scale $T\sim T_{\rm lc}$ is observed in X-ray emission from Galactic sources powered by black holes with masses $M_{\rm BH}\sim 10M_\odot$ \citep{remillard06}. Variability at the time scale $T_{\rm var}\sim T_{\rm lc}$ is observed also in \gr\ emission from blazars, a special type of AGN with jets aligned along the line of sight \citep{mrk421,m87,m87_theory,mrk501,pks2155,neronov09}. 

%%%%%%%%%%%%%%%%%%%%%%%%%%%%%%%%%%%%%%%%%%%%%%%
\begin{figure}
\begin{center}
\includegraphics[width=\linewidth]{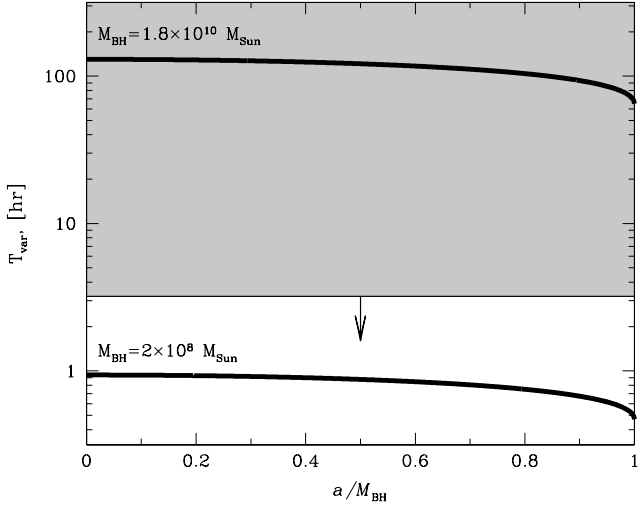}
\caption{Comparison of the upper bound on variability time scale of \OJ\ flares (grey shaded region) with the light crossing times of the two supermassive black holes in the system.}
\label{fig:OJ_jet_size}
\end{center}
\end{figure}
%%%%%%%%%%%%%%%%%%%%%%%%%%%%%%%%%%%%%%%%%%%%%%%

Fig. \ref{fig:OJ_jet_size} shows a comparison of the upper limit $T_{\rm var}\le 3.2$~hr on the variability time scale of the flares from \OJ\ with the light crossing time of the two supermassive black holes in the \OJ\ system. One can see that the upper bound is much lower than the light crossing time of the more massive black hole with $M_{\rm BH}\simeq 1.8\times 10^{10}M_\odot$, independently of the black hole rotation moment $a$. At the same time, the limit on the variability time scale is larger than the light crossing time of the smaller black hole. This implies that the observed \gr\ emission is generated by the jet emitted by the lower mass companion black hole in the system. 

Note, that the above result does not depend on the assumptions about relativistic motion of the \gr\ emission region. Indeed, the minimal possible time scale $\Delta T_{min}$ of emission from relativistic jet moving with bulk Lorentz factor $\Gamma$ does not depend on $\Gamma$ and is instead determined by the size of the non-moving "central engine" which ejected the jet \citep{celotti98,neronov09,neronov09a}. In the jet comoving frame, the minimal variability time scale $\Delta T_{min}'\sim \Gamma\Delta T_{min}$ is the light-crossing time of the smallest size \gr\ emitting "blobs", $\Delta T_{min}'\sim R_{min}'/c$.  \gr\ emitting blobs which were ejected by the jet's "central engine" of the size $R_{\rm CE}$ could not have  size smaller than $R_{min}'\ge R_{\rm CE}\Gamma$ in the comoving frame. This implies that $\Delta T_{min}\sim \Delta T_{min}'/\Gamma\sim R_{min}'/c\Gamma\ge R_{\rm CE}/c$, independently of $\Gamma$.

At the first sight, such conclusion looks counter-intuitive. Indeed, since the two black holes accrete matter from the same reservoir, the accretion rate onto more massive black hole should be much higher. In general, higher accretion rate should lead to production of a more powerful jet. 

However, the above argument does not take into account possible relativistic beaming effects, which are normally very significant in BL Lacs. \gr\ flux from relativistically beamed jet moving with bulk Lorentz factor $\Gamma$ at an angle $\theta$ with respect to the line of sight is boosted by a factor $\delta^4$, where $\delta=(\Gamma(1-\beta\cos\theta))^{-1}$ is the Doppler factor and $\beta$ is the bulk velocity of the jet.  A constraint on the bulk Lorentz factor of the jet produced by the $M_{\rm BH}\simeq 1.3\times 10^8M_\odot$ black hole could be found under assumption that the observed X-ray emission from the system \citep{suzaku} originates from the same jet region as the GeV emission.

The highest energy of \gr\ photons coming from the source is $E_{\gamma,max}\ge 30$~GeV (see Fig. \ref{fig:lcurve_zoom}). \gr s of such energies can produce $e^+e^-$ pairs in interactions with X-ray photons of the energies 
\begin{equation}
E_X\ge \frac{\Gamma^2m_e^2c^4}{(1+z)^2E_\gamma}\simeq 0.5\left[\frac{\Gamma}{4}\right]^2\left[\frac{E_{\gamma,max}}{30\mbox{ GeV}}\right]^{-1}\mbox{ keV}
\end{equation}
where we have assumed that typical collision angles for photons emitted from the jet are $\alpha\simeq \Gamma^{-1}$. Apparent (relativistically beamed) luminosity of \OJ\ in the soft X-ray band is $L_{\rm app}\simeq 3\times 10^{44}$~erg/s, which corresponds to the observed flux $F_X\simeq 10^{-12}$~erg/cm$^2$s \citep{suzaku}. Calculating the optical depth of \gr\ emission region of comoving size $R'\simeq\delta cT_{\rm var}/(1+z)$ and luminosity $L'\simeq \delta^{-4}(1+z)^2L_X$ one finds
\begin{equation}
\tau_{\gamma\gamma}\simeq 0.5\left[\frac{\delta}{4}\right]^{-6}\left[\frac{L_X}{10^{44}\mbox{ erg/s}}\right]\left[\frac{T_{\rm var}}{3.2\mbox{ hr}}\right]^{-1}
\end{equation} 
High energy \gr s could escape from the source if $\tau_{\gamma\gamma}< 1$. This condition imposes a restriction on the Doppler factor $\delta\gtrsim 4$.

Thus, the observed \gr\ flux from the jet is Doppler boosted by at least a factor $\delta^4\gtrsim 3\times 10^2$. It is interesting to note that even if emission from the higher mass black hole is not relativistically beamed toward observer on the Earth, it might be noticed in the spectrum of the source. Indeed,  assuming a simple Eddington-like scaling of the accretion rate and jet luminosity with the black hole mass, $L_i\sim M_{{\rm BH}i}$, one finds that the relativistically beamed luminosity of the jet from the lighter black hole $L\sim \delta^4L_1\ge 10^2L_1$ could, in fact, be comparable to the overall luminosity of the heavier  black hole, $L_2\sim (M_{\rm BH2}/M_{\rm BH1})L_1\simeq 1.4\times 10^2 L_1$. Fast variability of emission could therefore, serve as a tool for identification of the contribution of emission from the lighter black hole in the overall source spectrum. 

A common feature of all models of jet production by black holes is that matter ejection into the jet is associated with rotation of matter around the black hole and/or with rotation of the black hole \citep{blandford-znajek,blandford-payne}. This implies that characteristic time scale at which the properties of the jet could change is given by the period of rotation of the black hole itself or of the accretion flow onto the black hole. Period of rotation around a circular orbit at a distance $r$ from the black hole is given by \citep{bardeen}
\begin{equation}
\label{eq:P}
P(r)=2\pi(1+z)\frac{r^{3/2}\pm aR_g^{1/2}}{cR_g^{1/2}}\;,
\end{equation}
The $+$ ($-$) sign corresponds to the prograde (retrograde) orbit.  Stable circular
orbits exist only down to certain distance $r_{\rm ms}$ from the
BH. The period of rotation along the last prograde stable orbit at the
distance $r_{\rm ms}$ is
\begin{equation}
\label{eq:period}
P(r_{\rm ms})\simeq\left\{
\begin{array}{ll}
3\left[\frac{\displaystyle M_{\rm BH}}{\displaystyle 1.3\times 10^{8}M_\odot}\right]
\mbox{ hr,}&a=R_g\\
22
\left[\frac{\displaystyle M_{\rm BH}}{\displaystyle 1.3\times 10^{8}M_\odot}\right]\mbox{ hr,}&a=0
\end{array}
\right.
\end{equation}

Upper bound on the variability time scale $T_{\rm var}\le 3.2$~hr is much shorter than period of rotation around the non-rotating black hole and is comparable or smaller than the period of rotation around maximally rotating black hole with $a=R_g$. This means that  relativistic ejections into the jet, responsible for the observed flares, are produced by the matter moving in the direct vicinity of the black hole horizon, well inside the $R=6R_g$ radius of the last stable orbit around non-rotating black hole.  

%%%%%%%%%%%%%%%%%%%%%%%%%%%%%%%%%%%%%%%%%%%%%%
\section{Conclusions}
\label{CONCL}
%%%%%%%%%%%%%%%%%%%%%%%%%%%%%%%%%%%%%%%%%%%%%%

To summarize, we find that observations of \OJ\ in the $E>0.1$~geV energy band constrain the minimal timescale of flux variations of the source to be shorter than 3.2 hr. The upper limit on the minimal variability timescale imposes a restriction on the size of the jet formation region. We find that the size of the jet formation region in the \OJ\ system is much smaller than the size of the horizon of the more massive black hole in the binary black hole system powering the source. This means that the observed \gr\ emission is produced by the jet ejected from the smaller mass black hole. Higher apparent luminosity of the smaller mass companion is explained by the effect of relativistic beaming of \gr\ emission. Combining X-ray and \gr\ data, we find a restriction on the Doppler factor of the \gr\ emitting part of the jet, $\delta\gtrsim 4$. The observed variability time scale indicates that relativistic jet from the smaller mass black hole is formed close to the black hole horizon, well inside the last stable orbit around non-rotating supermassive black hole.  

\gr\ data provide a new insight in the physical model of the unique binary supermassive black hole system in  \OJ. \gr\ flaring activity is produced in connection with the passage of the smaller mass black hole through the accretion disk around the larger mass black hole during the periods of  close approach of the two black holes in the periastron of the binary orbit. Interaction of the smaller mass black hole with the larger mass black hole accretion disk leads to the transient episodes of ejection into relativistic jet from the smaller mass black hole. It appears that the  transient jet from the small mass black hole happens to be aligned along the line of sight, the fact responsible for the BL Lac type appearance of the source. It is not clear {\it a priori} if the jet from the smaller mass black hole forms only during the periastron passage or it exists throughout the binary orbit. If the jet is powered by the transient accretion onto the black hole, it would be natural to expect that the the small mass black hole jet (and the associated \gr\ emission) should disappear soon after the periastron passage on characteristic time scale of accretion. Otherwise, if the jet is powered by the rotation energy stored in the small mass black hole, it is natural to expect that the jet and the \gr\ from the jet would be persistent throughout the binary orbit. Systematic monitoring of the source evolution in \gr s on the orbital (11.7 years) time scale, which is now possible with {\it Fermi}, might clarify this question. It is interesting to note that if the jet from the smaller mass black hole is directed along the black hole spin axis, its alignment with the line of sight might be destroyed by the precession of the black hole spin axis (see \citet{valtonen06} for detailed discussion of the orbital evolution of the system). This would mean that the BL Lac appearance of of the source might be time-dependent. An immediate consequence of the mis-alignment of the smaller black hole jet with the line of sight should be the loss of the strong Doppler boosting of the flux. In the absence of Doppler boosting, the emission from the smaller mass black hole might become sub-dominant compared to the emission from the larger mass black hole. Study of the details of the overall time evolution and of the short time scale variability properties of the source along the orbit and from periastron to periastron should clarify the transient/permanent nature of the BL Lac appearance of the source.

\label{lastpage}


\begin{thebibliography}{99}

\bibitem[Aharonian et al.(2006)]{m87} Aharonian F. et al., 2006, Science, 314, 1424. 
\bibitem[Aharonian et al.(2007)]{pks2155} Aharonian F.A. et al., 2007, Ap.J., 664, L71.
\bibitem[Albert et al.(2007)]{mrk501} Albert J., et al. 2007, Ap.J., 669, 862.
\bibitem[Bardeen et al.(1972)]{bardeen} Bardeen, J.M., Press, W.H., 
Teukolsky, S.A., 1972, Ap.J. 178, 347. 
\bibitem[Blandford \& Payne(1982)]{blandford-payne} Blandford R. D., Payne D. G., 1982, MNRAS, 199, 883
\bibitem[Blandford \&Znajek(1977)]{blandford-znajek} Blandford R. D., Znajek R. L., 1977, MNRAS, 179, 433
\bibitem[Celotti et al.(1998)]{celotti98} Celotti A.L., Fabian A., Rees M., 1998, MNRAS,  293, 239
\bibitem[Ciprini et al.(2009)]{fermi_atel}  Ciprini S. et al., 2009, ATEL, 2256.
\bibitem[D'Ammando et al.(2009)]{swift} D'Ammando F., et al., 2009, ATEL, 2267.
\bibitem[Gaidos(1996)]{mrk421} Gaidos, J. A, 1996, Nature, 383, 319.

\bibitem[Harris \&Krawczynski(2006)]{jet_review} Harris D. E.; Krawczynski H., ARA\&A, 2006, 44, 463.
\bibitem[Lehto \& Valtonen(1996)]{lehto96} Lehto H.J., Valtonen M.J., 1996, Ap.J., 460, 207

\bibitem[Neronov \& Aharonian(2007)]{m87_theory} Neronov A. \& Aharonian F., 2007, Ap.J., 671, 85; 

\bibitem[Neronov et al.(2008)]{neronov09} Neronov A., Semikoz D., Sibiryakov S., 2008, MNRAS, 391, 949.

\bibitem[Neronov et al.(2008a)]{neronov09a} Neronov A., Semikoz D., Sibiryakov S., 2008a, 
AIP Conference Proceedings, 1085, 545.
\bibitem[Remillard \& McClintock(2006)]{remillard06} Remillard R.A., McClintock J.E., 2006, ARA\&A, 44, 49.
 
\bibitem[Seta et al.(2009)]{suzaku} Seta H., et al., 2009, PASJ, 61, 1011.

\bibitem[Sillanp\"a\"a et al.(1988)]{sillanpaa88} Sillanp\"a\"a A., Haarala, S., Valtonen, M. J., Sundelius, B., Byrd, G. G., 1988, Ap.J., 325, 628.
\bibitem[Stickel et al.(1989)]{stickel89} Stickel M., Fried, J. W., Kuehr, H., 1989, A\&AS, 80, 103.
\bibitem[Valtonen et al.(2006)]{valtonen06} Valtonen M.J. et al., 2006, Ap.J. 646, 36.
\bibitem[Valtonen et al.(2009)]{valtonen09}  Valtonen M.J. et al., 2009, Ap.J., 698, 781.
\bibitem[Valtonen et al.(2008)]{OJ_nature} Valtonen M.J. et al., 2008, Nature, 452, 851.
\bibitem[Zdziarski et al.(2010)]{zdz10} Zdziarski A., Neronov A., Cheryakova M., MNRAS, 2010, 403, 1873
\end{thebibliography}
\end{document}